\documentclass[a4paper]{article}
\usepackage[dvips]{graphicx}
\usepackage{color}

\begin{document}

\title {\bf Angular dependenceof the\\magnetisation AC loss:\\coated conductors, Roebel cables\\and double pancake coils}
\markboth{Angular dependence ...}{E. Pardo and F. Grilli}

\author{Enric Pardo$^{1,*}$ and Francesco Grilli$^2$\vspace{3 mm}\\
$^1$Institute of Electrical Engineering, Slovak Academy of Sciences,\\Bratislava, Slovakia\\ 
$^2$Karlsruhe Institute of Technology, Karlsruhe, Germany\\
$^*$ enric.pardo@savba.sk}

\date{}

\maketitle

\begin{abstract}
The AC loss in ReBCO coated conductors is large in situations when the conductors are subjected to a considerable magnetic field, like in rotating machines, transformers and high-field magnets. Roebel cables can reduce the AC loss in these cases. However, computer simulations are needed to interpret the experiments, understand the loss mechanisms, reduce the AC loss by optimising the Roebel cable and design the cryogenic system. In this article, we simulate and discuss the AC loss due to an applied magnetic field making an arbitrary angle with the cable and taking into account a realistic anisotropic field dependence of the critical current density. We study the AC loss in the superconductor parts for the limits of very high coupling currents and completely uncoupled strands. The simulations for the uncoupled case also describe a double pancake coil with no transport current. For the simulations, we use two different numerical methods with complementary strengths. This serves as a mutual check of the correctness of the simulation results, which agree to each other. Opposite than expected, we found that the AC loss does not only depend on the perpendicular component of the applied magnetic field. We also found that the AC loss for applied fields with an orientation below $7$ degrees with the strands surface is reduced more than one order of magnitude comparing to an untransposed cable. Therefore, we recommend to use Roebel cables for windings with important parallel components of the magnetic field, such as transformers and high-field magnets.
\end{abstract}






\section{Introduction}

ReBCO coated conductors present several advantages compared to other superconductors, such as their higher operation temperature (up to the liquid nitrogen temperature) and their good performance in large magnetic fields. Nowadays, coated conductors are produced in long lengths with uniform properties, so they are ready to be used in market applications. In particular, power applications are very promising, such as fault-current limiters, cables, rotating machines (including wind generators) and transformers. However, the AC loss in rotating machines and transformers is too high because of two reasons. First, the component of the AC magnetic field perpendicular to the tape is large. Second, these applications usually require high-current conductors made of several coated conductors. This results in thick cables with a substantial AC loss contribution from the parallel AC magnetic field. A similar situation arises in high-field \cite{weijers10IES} or large-scale DC magnets (such as for particle accelerators \cite{turenne10IPAC} or fusion). In this case, the it is still desirable to reduce the AC loss because it limits the ramp rate and generates significant histeresis loss, especially for liquid-helium cooled magnets. Roebel cables \cite{J:2006:Goldacker06,J:2009:Goldacker09,leghissa06AST,J:2008:Long08a,J:2009:LeeJK09a} reduce the ac loss caused by both components of an external magnetic field \cite{J:2009:Goldacker09,leghissa06AST,xieYY09IES}. In spite of the progress in Roebel cables, many aspects of their AC loss remain unknown \cite{turenne10IPAC}. A similar situation exists for pancake coils, widely used in windings.

Computer simulations are needed to interpret the experiments, understand the loss mechanisms, reduce the AC loss by optimising the Roebel cable or pancake coil and design the cryogenic system. However, to the best of our knowledge, there are no published simulations for the angular dependence of the magnetisation AC loss in Roebel cables and pancake coils.

The published simulations for the angular dependence of the magnetisation AC loss are only for a single tape or stacks of tapes. For a single tape, the published results are for a constant critical current density \cite{mikitik04PRBa}, $J_c$, or an anisotropic magnetic field-dependent $J_c$ \cite{ichiki04PhC,brandt05PRBa,enomoto05IES,stavrev05SST,EUCAS10fmsc}. However, the anisotropies and field\footnote{In order to avoid unnecessary repetition, we refer to the magnetic field as simply ``field" except when the electric and magnetic fields could be confused with each other.} dependences utilized in these articles are not realistic, since they take into account only a generic dependence \cite{brandt05PRBa}, a dependence extracted from data at high applied fields ($\ge$0.5 T) \cite{ichiki04PhC,kiuchi03PhC}, or without the correction for the self-field \cite{enomoto05IES,stavrev05SST}. Moreover, \cite{mikitik04PRBa,brandt05PRBa} neither calculate the complete AC cycle nor the AC loss. The work in \cite{EUCAS10fmsc} was for only two angles (90$^{\rm o}$ and 15$^{\rm o}$). For stacks, there is only one work for the coupled case (allowing the magnetisation currents to close in different tapes) and relatively high applied field amplitudes (50 mT)~\cite{stavrev05SST}. In addition, all these simulations were done assuming either a simplified anisotropy of $J_c$ by an elliptical dependence or assuming an isotropic superconductor. In contrast, actual coated conductors present complex anisotropies \cite{selvamanickam09PhC,holesinger09SST,zhangY09PhC,CoatedIc}.

Simulations on Roebel cables are only for perpendicular applied fields \cite{roebelmeassim,roebelcomp} or transport currents \cite{roebelmeassim,roebelcomp,jiangZ11SST}. For double (or single) pancake coils (or stacks of tapes) the magnetisation AC loss of stacks of tapes was simulated only for perpendicular applied fields in \cite{tapesfull,grilli06PhC,roebelmeassim,yuanW10JAP,roebelcomp,prigozhin11SST}.

The only measurements on the angular dependence of the AC loss are in \cite{terzieva11SST}. However, there exists extensive work on coated conductors \cite{amemiya04SSTa,ogawa05Cry} and stacks of them \cite{iwakuma04PhC,iwakuma05IES,jiangZ08SST}. All these measurements for single tapes and stacks were for coated conductors at an early stage of development, with poor artificial pinning and anisotropies approximately elliptical. As a consequence, they may not be representative for the present nano-engineered material \cite{selvamanickam09PhC,holesinger09SST,zhangY09PhC}. For Roebel cables there are also measurements for perpendicular applied fields \cite{roebelmeassim,lakshmi10SSTa,J:2009:LeeJK09a,terzieva11SST}, parallel ones \cite{J:2010:lakshmi10b} and transport currents \cite{roebelmeassim,J:2010:jiangZ10}. 

In this article, we simulate and discuss the AC loss due to an applied magnetic field at an arbitrary angle with the cable and we take into account a realistic anisotropic field dependence of the critical current density. We study the AC loss in the superconductor parts for the limits of very high coupling currents (coupled case) and completely uncoupled strands (uncoupled case). The 2D simulations for the uncoupled case also describe a double pancake coil with no transport current. Therefore, the results and discussions for the uncoupled case are also valid for double pancake coils. Additionally, we also discuss the details of the AC loss in a single tape. For the simulations, we use two different numerical methods with complementary strengths: the Minimum Magnetic Energy Variation (MMEV) and a Finite Element Method (FEM) with the $H$-formulation and edge elements (see section \ref{s.simmethods}). This serves as a mutual check of the correctness of the simulation results. Moreover, it is also a check of the applicability of the sharp $E(J)$ relation from the critical-state model ($E$ and $J$ are the electric field and the current density, respectively), because MMEV assumes this sharp $E(J)$ relation and FEM, a smooth one.

This article is structured as follows. In section \ref{s.models} we outline the simulation models and the anisotropic field dependence of $J_c$ for the calculations. In section \ref{s.results} we present and discuss the results for a single tape and a Roebel cable in the coupled and uncoupled cases. For the Roebel cable, we do not only present the AC loss but also the field and current distribution for some cases. Finally, in \ref {s.conclusions} we present our conclusions.


\section{Models}
\label{s.models}

In this section, we first outline both simulation methods, their complementary strengths and some technical details (section \ref{s.simmethods}). Afterwards, we detail the anisotropic field dependences of $J_c$ in the simulations: { two anisotropic (a realistic one for YBCO and an elliptical one) and one isotropic field dependences (section \ref{s.JcBth}).}


\subsection{Simulation methods}
\label{s.simmethods}

In this article, we use two different numerical methods to obtain the current distribution, the magnetic field and the AC loss, as we did in a previous work \cite{roebelcomp}. These models are the Minimum Magnetic Energy Variation (MMEV) method and the Finite Element Method (FEM) with $H$-formulation and edge elements. The comparison of their results serves as a mutual check of the correctness of the methods, as well as of the assumption of the critical-state model {for modelling high-temperature superconductors.}

These methods present different strengths. The MMEV method is generally faster than FEM \cite{roebelcomp}. This is an advantage to calculate situations with many tapes, such as windings \cite{pancaketheo,simHTS11}. Since MMEV is a user-programmed software (the program for this article is written in FORTRAN language), it is possible to control the processes, make further improvements and include it in larger programs. In addition, it is possible to exploit the vast {collection of} existing free-source numerical routines in FORTRAN or C++. The FEM model is more versatile because it uses a commercial software \cite{S:COMSOL}. In particular, it can simulate non-linear magnetic materials interacting with the superconductor~\cite{nguyen10SST} and multi-physics problems, as for example the coupling of electromagnetic and thermal effects~\cite{roy08IES}. Moreover, in contrast to MMEV, the FEM model can simulate over-critical currents because it takes into account a smooth $E(J)$ relation.

Both methods are 2D models, reducing the problem to solving the cross-section of the cable. In more detail, they assume that the transposition length of the cable is much larger than its thickness and width, so the cable is well approximated by a set of infinitely long tapes parallel to each other, as we detailed in \cite{roebelcomp}.


\subsubsection{The Minimum Magnetic Energy Variation (MMEV) method}
\label{s.mmev}

The MMEV method assumes the sharp $E(J)$ relation of the critical-state model, where $E$ is the electrical field. It is based on a variational principle proposed by Prigozhin \cite{prigozhin97IES}, which finds the current distribution by minimising the magnetic energy variation, and a fast non-standard minimisation routine. This routine has been developed incrementally in several articles. First, Sanchez and Navau solved a cylinder in an applied magnetic field under certain restrictions \cite{sanchez01PRB}. Later on, Pardo {\it et al.} developed the general method for tapes under any combination of applied magnetic field and transport current \cite{HacIacinphase}. The latest stage of the method is published in \cite{pancakeBi,pancakeFM,simHTS11}, where \cite{pancakeBi,pancakeFM} and \cite{pancakeFM} take into account the field dependence of $J_c$ and the interaction with linear magnetic materials, respectively. Recently, we applied the method to Roebel cables \cite{roebelmeassim,roebelcomp}, but using a constant $J_c$.


\subsubsection{The Finite Element Method (FEM) model}
\label{s.fem}

The FEM model assumes a smooth $E(J)$ relation, usually a power law $E(J)=E_c (J/J_c)^n$, where $E_c$ is the voltage-per-length criterion for the critical current density $J_c$ {(usually set equal to $\rm 10^{-4}~Vm^{-1}$)} and $n$ is the flux-creep exponent. In this work, we assume that $n$ is constant, neglecting the magnetic field dependence of this parameter. The state variable of the FEM model are the magnetic field components; as a consequence the implementation of the $J_c({\bf B})$ dependence is straightforward since the magnetic flux components are immediately available from the state variables by means of the ${\bf B}=\mu_0{\bf H}$ relation.
More details about the model implementation can be found in~\cite{brambilla07SST}.


\subsubsection{Tape and Roebel cable parameters}

The anisotropic field dependences of $J_c$ for the simulations are detailed in \ref{s.JcBth}. These dependences are based on measurements \cite{CoatedIc} of YBCO coated conductors from SuperPower, Inc. \cite{SuperPower}. The functions for $J_c$ describe the experimental anisotropic field dependence with several degrees of approximation: realistic, elliptic and isotropic. For all cases, we do not take into account the metal parts of the tapes in the calculations, ignoring the effects of eddy currents.

For the single tape, the dimensions for the simulations are 4.16 mm for the width and 1.4 $\mu$m for the thickness.

For the Roebel cable, we chose the geometry of a cable composed of 14 strands, which was manufactured at the Karlsruhe Institute of Technology~\cite{roebelmeassim}. The strands are 1.98~mm wide, their lateral separation is 200 $\rm \mu m$ and their vertical separation (i.e. the distance between the superconducting films) is 140~$\rm \mu m$. The total critical current of the cable at 77~K (determined with the 1~$\mu$V/cm criterion over a distance of 30~cm) is 465 A. In the simulations, the strands (from now on we call them ``tapes") have the following dimensions: width $w=1.98~\rm mm$ and thickness $d=1.4~\mu$m. For the power-law resistivity used in the FEM simulations, $E_c=10^{-4}~\rm V/m$ and $n=35$ (except stated otherwise). The frequency of the applied field is 100~Hz.


\subsubsection{Coupled and uncoupled cases}
\label{s.coupunc}

For the cable simulations, we distinguish between the coupled and uncoupled cases. The coupled case assumes that the resistance per transposition length between the tapes is very small, so the magnetisation current loops can close freely between any tape of the cable. The uncoupled case assumes a very large resistance per transposition length between the tapes, so the current loops must close within each tape. From the computation point of view, this means that for the coupled case there is only one current constrain (zero net current in the whole cable), while for the uncoupled one there are as many current constrains as tapes (zero net current in each tape).

A cable made of untransposed tapes corresponds to the coupled case because the tapes are interconnected together at the current leads. Indeed, Polak {\it et al.} \cite{polak09SST} experimentally found that for tape lengths of around 10 cm or above, interconnected tapes in parallel are already fully coupled.

A double pancake coil exactly corresponds to the uncoupled case. Then, all the results and discussions for the Roebel cable in the uncoupled case are also valid for double pancake coils.


\subsubsection{Discretisation of the superconducting domain}
\label{s.elements}

Due to their high aspect ratio, ReBCO superconductors are often approximated as 1-D objects, where the variation of the electromagnetic quantities along the thickness is neglected. However, in certain cases, this approximation is not correct. In particular, if we consider the cases shown in this paper, this happens for two reasons. First, for the uncoupled case (or a single tape) and low angles and low fields, the AC loss is dominated by the current penetration across the thickness of the tape (figures \ref{f.Jth07} and \ref{f.Qunc}). Second, for a field-dependent $J_c$ and the uncoupled case (or a single tape), taking only one element in the thickness (or a 1D approximation) neglects the influence  of the local parallel field on $J_c$. This is because the parallel component of the field is anti-symmetric with respect to the mid-plane of the tape, so both its value at the mid-plane and its average in the thickness vanish. Taking only one element in the thickness induces significant errors for the uncoupled case and single tapes at applied fields below the self-field when $J_c$ is not strongly anisotropic, as is the case of ReBCO coated conductors.

The simulations use the following number of elements. For the coupled case, there are 1 element in the thickness and between 100 and 500 elements in the width per tape, with larger values for lower applied magnetic fields. For the uncoupled case and single tapes, there are between 1 and 20 elements in the thickness and 100 and 500 elements in the width. Higher {numbers of} elements in the thickness and the width are for lower applied magnetic fields. For the particular case of 0$^{\rm o}$ (parallel applied field) and the uncoupled situation, we use 50 elements in the thickness and 10 in the width.


\subsection{Angular and field dependence}
\label{s.JcBth}

\begin{figure}[p]
\begin{center}
\includegraphics[width=12cm]{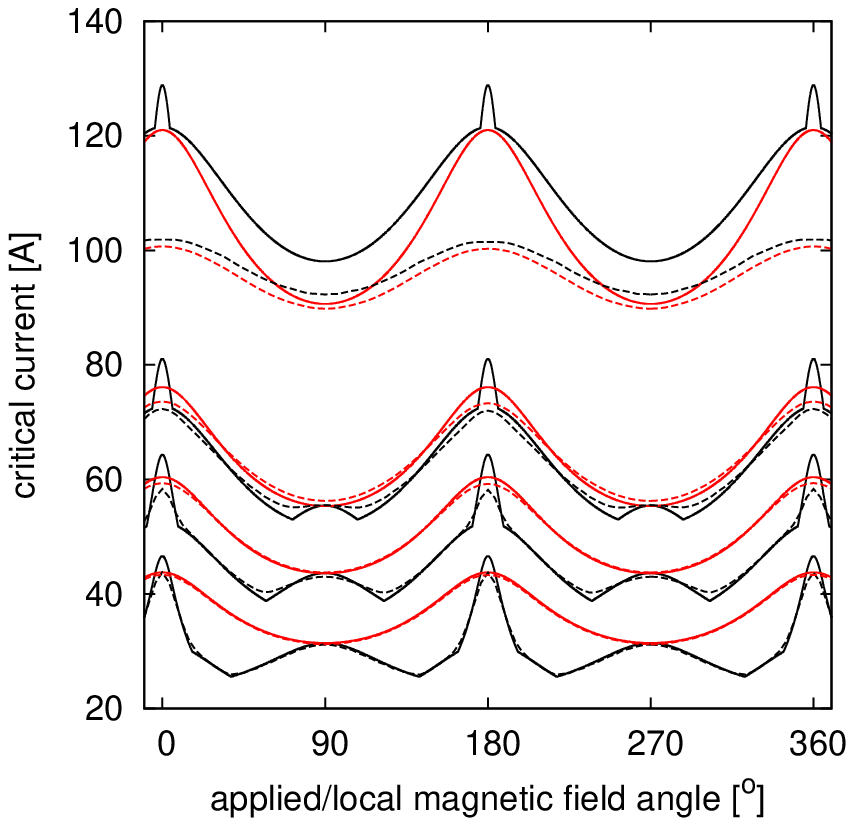}
\includegraphics[width=12cm]{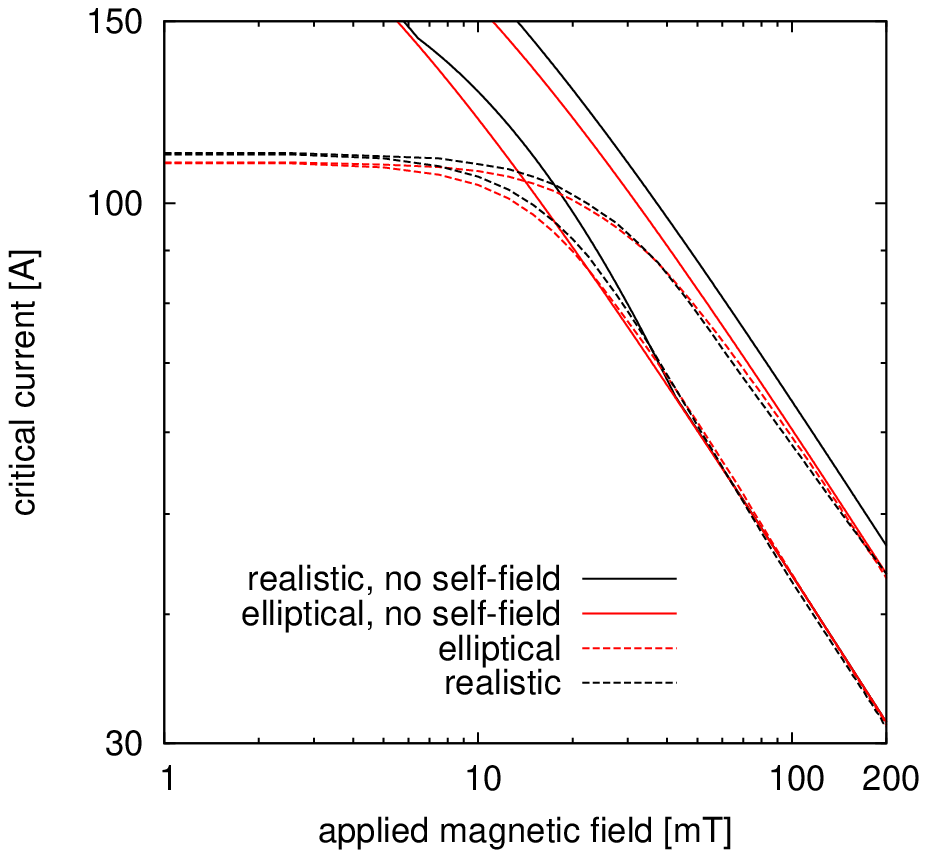}
\end{center}
\caption{In the top graph, the applied magnetic field is, from top to bottom: 20, 60, 100 and 200 mT. In the bottom graph, the lines are for 0$^{\rm o}$ and 90$^{\rm o}$, from top to bottom.}\label{f.Icth}
\end{figure}

In this article, we study the effect of three different field and angular dependences of the critical current density, $J_c(B,\theta)$: isotropic superconductor, elliptical anisotropy and a realistic field and angle dependence. This latter dependence presents a peak in both the $ab$ and $c$ directions and contains three different contributions with elliptical anisotropy (figure \ref{f.Icth}a). In the following sections, we study the error in the computations as a consequence of choosing a simplistic angular dependence, i.e. isotropic superconductor or elliptical anisotropy. In order to do so, we assume that the superconductor is perfectly described by the realistic description. Then, we simulate a measurement of the in-field critical current for an applied field in the $ab$ and $c$ directions (figure \ref{f.Icth}b). This is the data available in many experiments. With this data, we extract the parameters assuming an elliptical anisotropy. This simplification describes well the critical current in the $ab$ and $c$ directions but not for intermediate angles at large applied magnetic fields. If we simplify further, to an isotropic superconductor, we could only expect to fit the critical current at one orientation. Moreover, it is not possible to fit the field dependence at low applied fields and high applied fields at the same time because the self-field changes the orientation of the local magnetic field. In this article we choose to take the parameters that fit the critical current at high applied fields in the $ab$ direction.

\subsubsection{Realistic field and angle dependence}
\label{s.JcBthreal}

Many ReBCO coated conductor tapes present an angular dependence with a peak in the $ab$ and $c$ directions. Reference \cite{CoatedIc} found that an expression based on three contributions that describes a coated conductor {manufactured by SuperPower, Inc. \cite{SuperPower}}. Here, we use a simplified version of that field and angle dependence of $J_c$, which reproduces the main features of the experimental critical current in \cite{CoatedIc}.
\begin{equation}
\label{JcBth}
J_{c}(B,\theta)=\max\{J_{c,ab}[Bf_{ab}(\theta)],J_{c,c}[Bf_{c}(\theta)],J_{c,i}[Bf_i(\theta)]\}
\end{equation}
with
\begin{eqnarray}
J_{c,ab}(B) & = & \frac{J_{0p}}{(1+B/B_{0ab})^{\beta}}, \label{Jcab} \\
J_{c,c}(B) & = & \frac{J_{0p}}{(1+B/B_{0c})^{\beta}}, \label{Jcc} \\
J_{c,i}(B) & = & \frac{J_{0i}}{(1+B/B_{0i})^{\alpha}}, \label{Jci}
\end{eqnarray}
and
\begin{eqnarray}
f_{ab}(\theta) & = & \sqrt{\cos^2\theta+u_{ab}^2\sin^2\theta}, \label{fab0} \\
f_{c}(\theta) & = & \sqrt{u_c^2\cos^2\theta+\sin^2\theta}, \label{fc} \\
f_i(\theta) & = & \sqrt{\cos^2\theta+u_i^2\sin^2\theta}. \label{fi}
\end{eqnarray}
We take the following values for the parameters: $J_{0p}=4.9\cdot 10^{10}$~A/m$^2$, $J_{0i}=3.2\cdot 10^{10}$~A/m$^2$, $B_{0ab}=4.6$~mT, $B_{0c}=2.0$~mT, $B_{0i}=32$~mT, $\beta=0.48$, $\alpha=0.9$, $u_{ab}=8.3$, $u_c=1.8$ and $u_i=1.7$.

\subsubsection{Elliptical anisotropy}

Now, we consider the Blatter's \cite{blatter94RMP} scaling law for the field and angular dependence, also known as elliptical anisotropy:
\begin{equation}
\label{JcBthe}
J_{c}(B,\theta)=J_{c,e}[Bf_{e}(\theta)]
\end{equation}
with
\begin{eqnarray}
J_{c,e}(B) & = & \frac{J_{0e}}{(1+B/B_{0e})^{\beta}}, \label{Jce} \\
f_{e}(\theta) & = & \sqrt{\cos^2\theta+u_{e}^2\sin^2\theta}. \label{fe}
\end{eqnarray}
For this field and angle dependence we set the parameters in the following way. We take the parameters for which the critical current $I_c$ as a function of the applied magnetic field $B_a$ for the elliptical dependence, equations (\ref{JcBthe})-(\ref{fe}), fits the best to $I_c$ for the realistic dependence, equations (\ref{JcBth})-(\ref{fi}), if the applied field is in the $ab$ or $c$ directions (see figure \ref{f.Icth}b). For low applied magnetic fields, the critical current is smaller than the integration of $J_c(B_a,\theta)$ over the volume because the self-field increases the local magnetic field. The parameters for the elliptical anisotropy are $J_{0e}=4.602\cdot 10^{10}$A/m$^2$, $B_{0e}=4.6$mT, $u_e=2.015$ and $\beta$ takes the same value as for equations (\ref{Jcab}) and (\ref{Jcc}), $\beta=0.48$.

\subsubsection{Isotropic superconductor}

Finally, the simplest case is that of an isotropic superconductor, with no angular dependence of $J_c$. 
\begin{equation}
\label{Jciso}
J_{c,{\rm iso}}(B,\theta) = \frac{J_{0}}{(1+B/B_{0})^{\beta}}.
\end{equation}
For this field dependence, we take the same parameters as the field dependence for the elliptic case in the $ab$ direction, equation (\ref{Jcc}): $B_0=4.6$mT, $J_{0}=4.9\cdot 10^{10}$~A/m$^2$ and $\beta=0.48$ (if we wanted to take the dependence in the $c$ direction, the isotropic field dependence would be with $B_0=B_{0e}/u_e=2.283$mT).

With the isotropic dependence we do not find the parameters that fit the critical current for the realistic dependence. This is because an isotropic critical current density cannot describe the in-field critical current of a tape with an anisotropic superconductor, not even for a particular orientation.


\section{Results and discussion}
\label{s.results}

\subsection{AC loss for a single tape}
\label{s.tape}

\begin{figure}[p]
\begin{center}
\includegraphics[width=12cm]{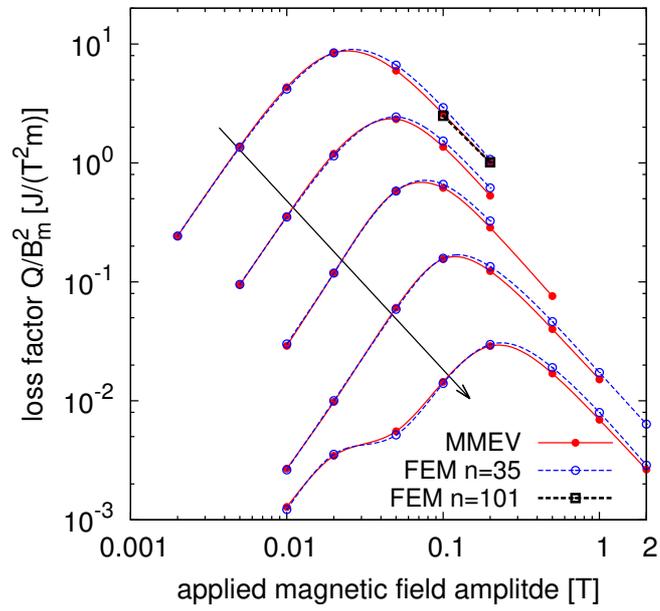}
\end{center}
\caption{AC loss per unit length and cycle, $Q$, for a single tape with the realistic anisotropic field dependence of $J_c$ in section \ref{s.JcBth}. The angle of the applied magnetic field is 90, 30, 15, 7, 3 degrees in the arrow direction.}\label{f.Qtape}
\end{figure}

\begin{figure}[p]
\begin{center}
\includegraphics[width=10cm]{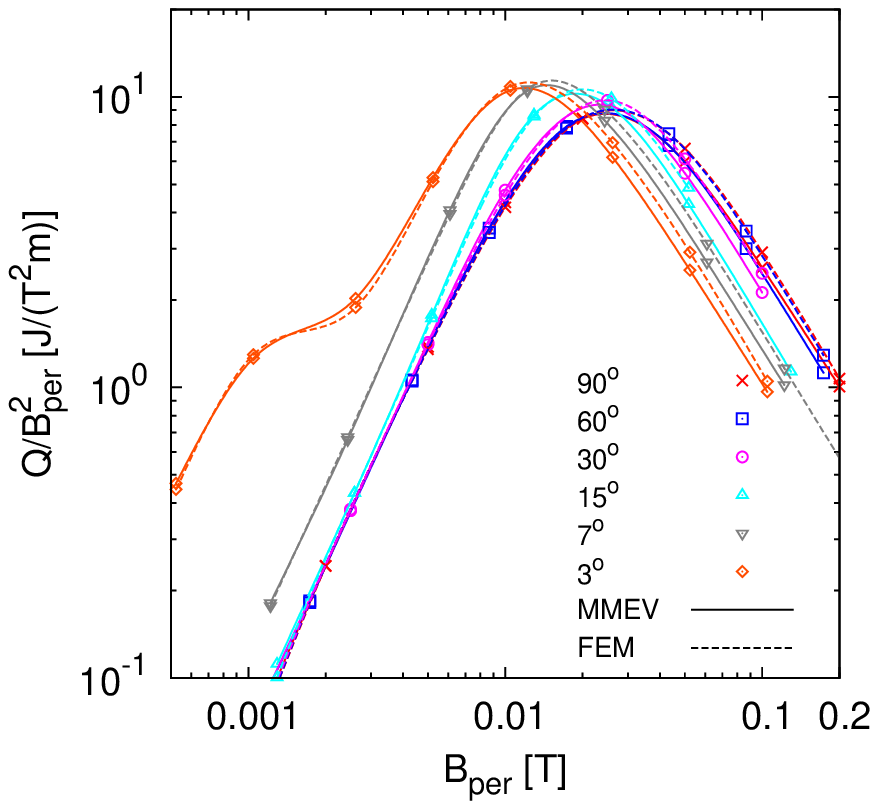}
\includegraphics[width=10cm]{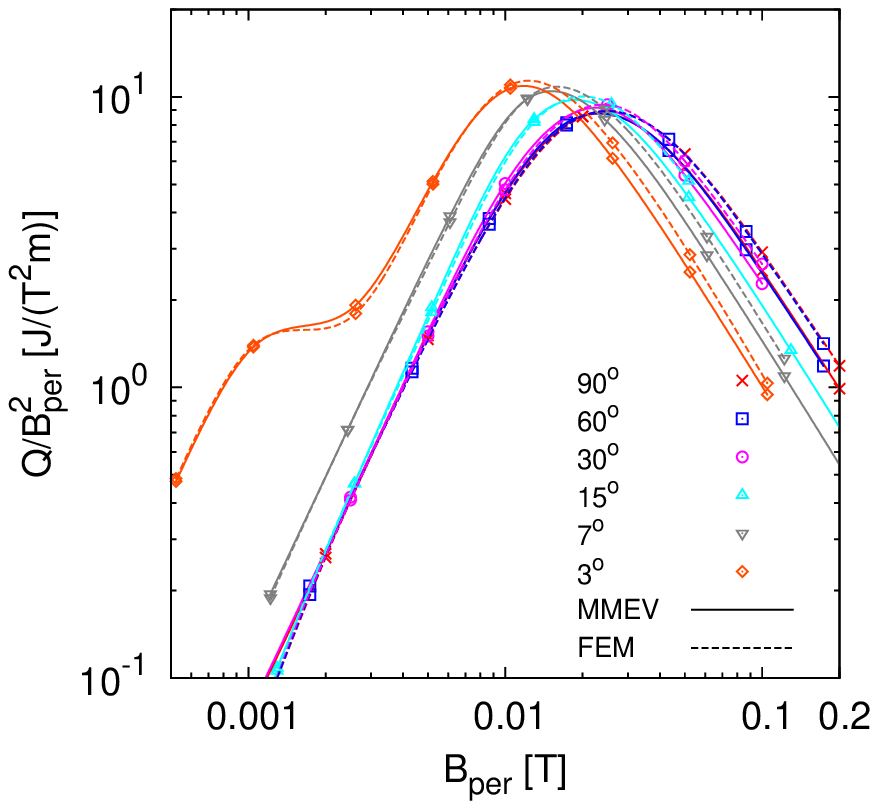}
\includegraphics[width=10cm]{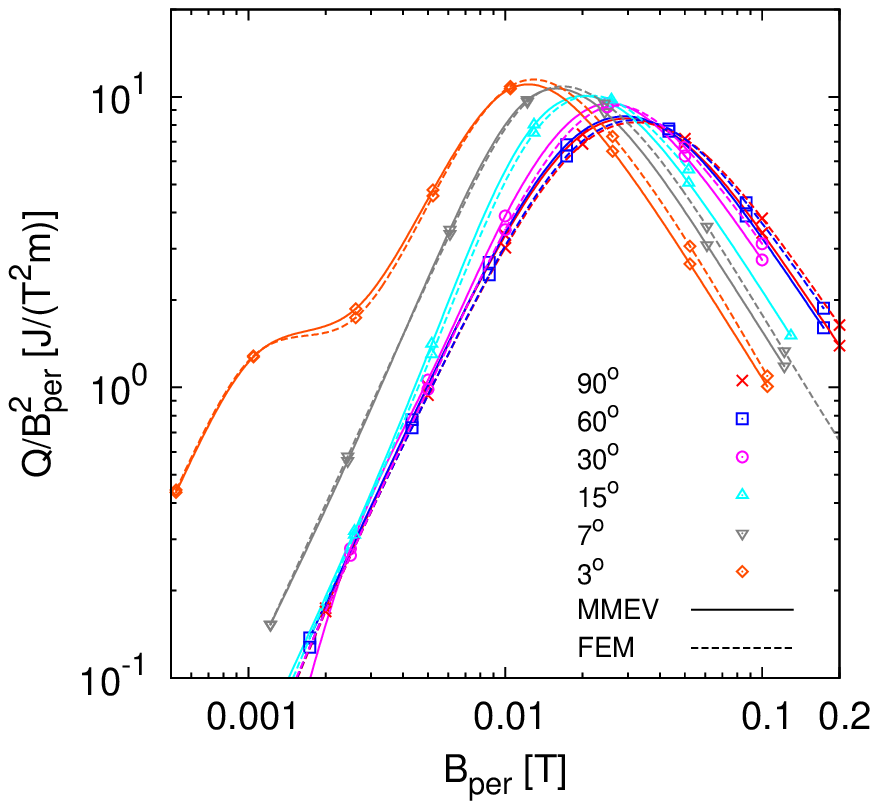}
\end{center}
\caption{Loss factor $Q/B_{\rm per}^2$ as a function of the perpendicular applied field amplitude $B_{\rm per}$ for a single tape for the anisotropic field dependences in section \ref{s.JcBth}: realistic, elliptic and isotropic (from top to bottom). The symbols distinguish different angles of the applied field and the solid/dash lines distinguish the results from MMEV and FEM.}\label{f.Qtapenorm}
\end{figure}

In this section we present and discuss the AC loss in a single tape for an applied magnetic field with several orientations and the three angular dependences of section \ref{s.JcBth} (results in figures \ref{f.Qtape},\ref{f.Qtapenorm}). Actually, we present the loss factor relative to the applied magnetic field amplitude, $Q/B_m^2$ ($Q$ is the loss per cycle and length and $B_m$ is the applied field amplitude) or to the component of the applied magnetic field perpendicular to the tape surface, $Q/B_{\rm per}^2$.

The results from both simulation methods agree (figures \ref{f.Qtape} and \ref{f.Qtapenorm}). However, there is a small discrepancy at large amplitudes because of the finite flux-creep exponent for the FEM simulations (exponent 35). Indeed, when {we increase} the flux creep exponent to 101, this discrepancy vanishes (figure \ref{f.Qtape}).

As expected, the AC loss decreases with decreasing the orientation angle \cite{oomen97APL,
ichiki04PhC,enomoto05IES,amemiya04SSTa,ogawa05Cry}. This is because the AC loss in a thin tape is dominated by the perpendicular component of the applied magnetic field and this component of the applied magnetic field decreases with decreasing the angle. Moreover, with decreasing the angle, the peak in the loss factor moves to higher applied fields because the perpendicular component of the applied field decreases.

The shoulder at low applied magnetic fields for low angles (curve for 3$^{\rm o}$ in figures \ref{f.Qtape} and \ref{f.Qtapenorm}) is because of the current penetration across the thickness \cite{EUCAS10fmsc}. When the perpendicular component of the applied field is low, the AC loss due to the parallel component becomes important.

The loss factor relative to the perpendicular applied magnetic field, $Q/B_{\rm per}^2$, is generally not independent on the angle (figure \ref{f.Qtapenorm}), in contrast to the conclusions of earlier work \cite{ichiki04PhC,enomoto05IES,amemiya04SSTa,ogawa05Cry,iwakuma04PhC,iwakuma05IES,jiangZ08SST}. However, there actually is this angle independence for large and moderate angles ($\ge$30$^{\rm o}$). Apart from the shoulder at low amplitudes, for low angles ($\le$15$^{\rm o}$) the peak of $Q/B_{\rm per}^2$ increases, shifts to lower fields and becomes slightly narrower. {If we compare} 3$^{\rm o}$ with 90$^{\rm o}$, the peak for 3$^{\rm o}$ is at around 3 times lower $B_{\rm per}$, resulting in around 3 times larger ac loss for $B_{\rm per}$ close to the peak for 3$^{\rm o}$. The difference in AC loss is similar for large $B_{\rm per}$. This angle dependence of $Q/B_{\rm per}^2$ is caused by the anisotropy and field dependence of the critical current density.

The results for the isotropic field dependence and the elliptical one reveal the different effects of the anisotropy and the field dependence, as follows. 

The increase of the peak in $Q/B_{\rm per}^2$, its shift to lower angles and its narrowing when decreasing the angle are due to the field dependence because these effects are already present for the isotropic $J_c$ (figure \ref{f.Qtapenorm}c). In particular, the shift of the peak of $Q/B_{\rm per}^2$ to lower $B_{\rm per}$ with decreasing the angle is because of the following. For the same $B_{\rm per}$, lower angles result in higher applied field magnitudes and, consequently, in a lower average $J_c$ in the AC cycle. Then, the tape saturates for lower $B_{\rm per}$ and the loss factor shifts accordingly. The increase in the peak of $Q/B_{\rm per}^2$ and its narrowing is due to the stronger field dependence of $J_c$ as a function of $B_{\rm per}$ for lower $B_{\rm per}$ \cite{chen91JAP}. Indeed, the field dependence of (\ref{Jciso}) as a function of $B_{\rm per}$ is $J_c={J_{0}}/{(1+B_{\rm per}/(B_{0}|\sin{\theta}|))^{\beta}}$, resulting in an effective field constant $B_{0,{\rm eff}}=B_0|\sin{\theta}|$ which decreases with the angle $\theta$ and, therefore, strengthens the field dependence with decreasing the angle.

An elliptical anisotropy weakens the angular dependence of $Q/B_{\rm per}^2$ with respect to the realistic or isotropic dependence (figure \ref{f.Qtapenorm}b). This is because the angular dependence in $J_c$ of (\ref{fe}) becomes $f_e(\theta)\approx u_e|\sin\theta|$ for large enough angles $\theta$, resulting in a $B_{\rm per}$ dependence independent on the angle, $J_c=J_{0e}/(1+B_{\rm per}u_e/B_{0e})^\beta$. The angle range {where} $J_c$ becomes independent of $B_{\rm per}$ becomes wider for larger anisotropies, that is larger $u_e$. This explains the published angle independence of $Q/B_{\rm per}^2$ as a function of $B_{\rm per}$ for Bi2223, with a large anisotropy \cite{oomen97APL}.

The AC loss for the realistic case presents the main features of the isotropic case regarding the position, height and width of the peak. This is because the anisotropy of $J_c$ is not very strong. The main difference from the isotropic case is that the angular dependence of $Q/B_{\rm per}^2$ loss is smaller, especially for $B_{\rm per}$ below the peak. This is because for low fields, the realistic $J_c$ approaches an elliptical dependence (figure \ref{f.Icth}a), therefore a weaker angular dependence of $Q/B_{\rm per}^2$.

\subsection{Magnetic field and current distribution in a Roebel cable}

\begin{figure}[p]
\begin{center}
\includegraphics[width=12cm]{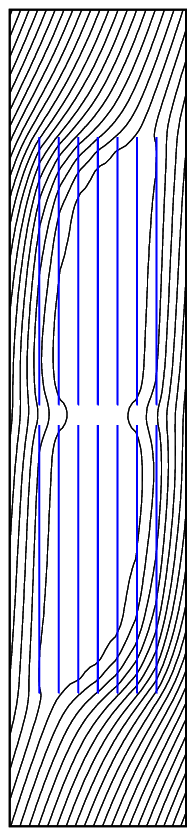}
\includegraphics[width=12cm]{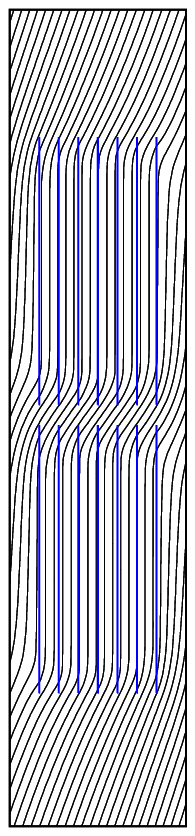}
\end{center}
\caption{The magnetic flux lines for the coupled case (up) are very different from the uncoupled one (bottom). These flux lines are calculated with the MMEV model at the peak of the AC cycle for $\theta$=15$^{\rm o}$ and $B_m$=50 mT.}\label{f.Blines}
\end{figure}

The magnetic field distribution for the coupled and uncoupled cases are very different (figure \ref{f.Blines}). For the coupled case, the superconductor shields the applied magnetic field as much as possible in all the cable volume. For the uncoupled case, the superconductor only shields the volume of each tape individually. This is in accordance to the published results for perpendicular applied magnetic fields \cite{tapesfull,roebelcomp}. In addition, for the uncoupled case, the superconductor shields the perpendicular component of the applied magnetic field within each stack of tapes but not in the gap between the stacks, where it concentrates. As a result, the angle of the magnetic field in this gap increases. Another consequence is that the magnetic field within each stack is parallel to the tape surfaces. Moreover, the magnetic field in between the tapes of the stack is uniform. Actually, this is always the case when the magnetic field in the horizontal separation between tapes is parallel to their surface \cite{roebelcomp}.

\begin{figure}[p]
\begin{center}
\includegraphics[width=10cm]{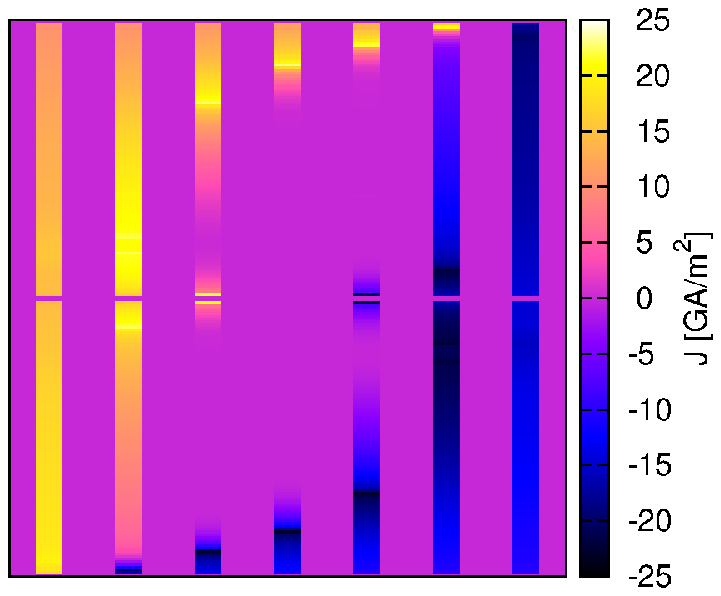}
\includegraphics[width=10cm]{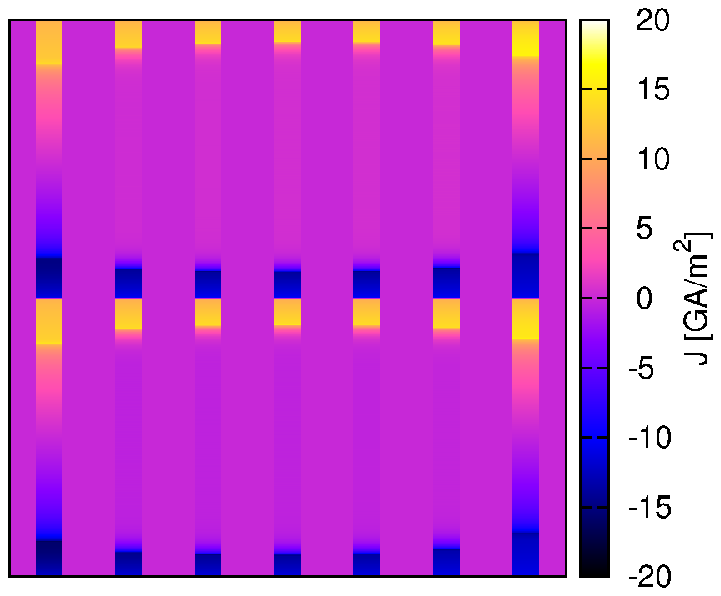}
\end{center}
\caption{This cross-section shows that the current distribution for the coupled (up) and uncoupled (bottom) cases are evidently different (the current distribution is calculated with the MMEV model). The situation is for the peak of the AC cycle for an applied field of $\theta$=15$^{\rm o}$ and 50 mT of amplitude. This current distribution is for the 1D approximation: only one element in the superconductor thickness. For better visualisation, the horizontal and vertical axis are not on scale and the represented tape thickness in not in scale with the horizontal separation between tapes.}\label{f.Jth15}
\end{figure}

\begin{figure}[p]
\begin{center}
\includegraphics[width=10cm]{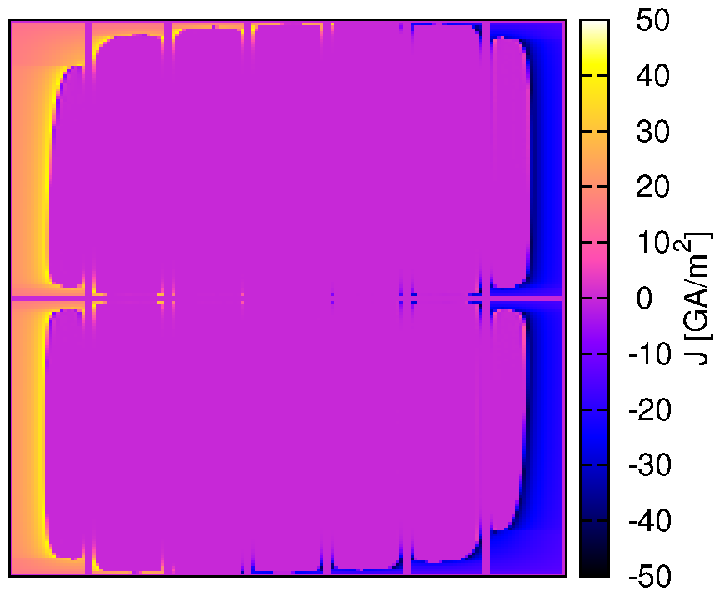}
\includegraphics[width=10cm]{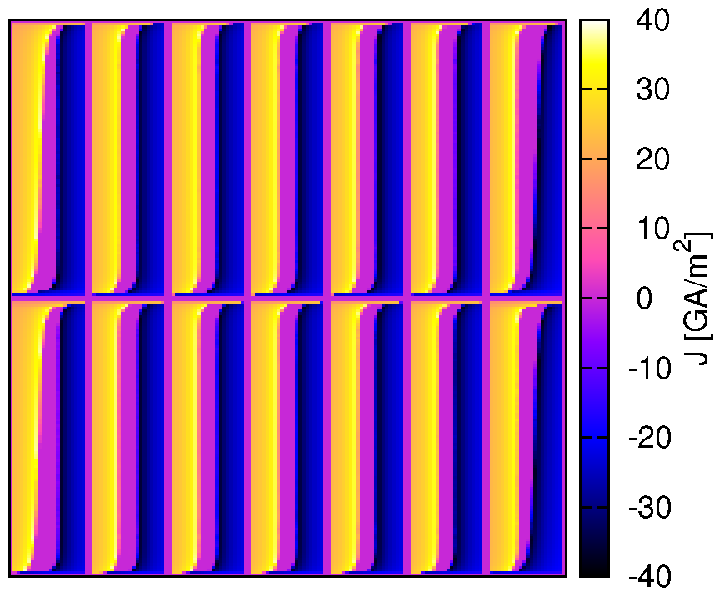}
\end{center}
\caption{This cross-section shows that the current distribution for the coupled (up) and uncoupled (bottom) cases are evidently different (the current distribution is calculated with the MMEV model). The situation is for the peak of the AC cycle for an applied field of $\theta$=7$^{\rm o}$ and 20 mT of amplitude. This current distribution calculated with 20 elements in the thickness shows the details of the current penetration across the tapes thickness. For better visualisation, the horizontal and vertical axis are not on scale and the represented tape thickness is not in scale with the horizontal separation between tapes.}\label{f.Jth07}
\end{figure}

In some cases it is not enough to take only one element in the thickness of the tape (1D approximation), see section \ref{s.elements}. This is the case for the uncoupled situation and low applied magnetic fields. An example is for an applied magnetic field at 7$^{\rm o}$ and 20 mT of amplitude, figure \ref{f.Jth07}, where the AC loss is dominated by the current penetration across the thickness of the tape (figure \ref{f.Qunc}). On the contrary, for an applied field of, for example, 15$^{\rm o}$ and 50 mT of amplitude (figure \ref{f.Jth15}), one element in the thickness is sufficient for AC loss and magnetic field calculations. This is because the main AC loss contribution is from the penetration across the width of the tape and the average parallel magnetic field in the tapes thickness is nonzero.

As expected, the current distribution between the coupled and uncoupled cases is very different (figures \ref{f.Jth15} and \ref{f.Jth07}). For the coupled case, the current penetrates roughly as in a monoblock. With an oblique applied magnetic field, the current penetrates faster from the top-left and bottom-right corners, figures \ref{f.Jth15}a and \ref{f.Jth07}a. This is consistent with the calculations for one single tape \cite{mikitik04PRBa} and the field penetration in figure \ref{f.Blines}a. For the uncoupled case, the net current in each tape is zero with qualitatively similar current penetration in each tape, except in the tapes at the boundaries of the cable (figures \ref{f.Jth15}b and \ref{f.Jth07}b). The only similarity between the coupled and uncoupled cases is that the current distribution is anti-symmetric with respect to the central point of the tape. This is because of the geometry of the cable and the applied magnetic field. 

The current distribution for the 1D approximation allows an accurate quantitative comparison between the simulation methods (MMEV and FEM), rather than a qualitative comparison between current distributions from colormaps like figures \ref{f.Jth15} and \ref{f.Jth07} or the comparison of the AC loss in a logarithmic scale, such as in figures \ref{f.Qtape},\ref{f.Qtapenorm},\ref{f.Qcoup} and \ref{f.Qunc}. The simulation methods agree very well, better than for a constant $J_c$ \cite{roebelcomp}.  Moreover, the effect of the smooth $E(J)$ relation for the FEM simulations is not very important for low applied field amplitudes \cite{roebelcomp}. With increasing the amplitude, the local electro-motive force due to the applied magnetic field increases, and so does the local electric field and $J$ for a smooth $E(J)$ relation. This is indirectly seen in the higher AC loss at high amplitudes for the FEM model (figures \ref{f.Qcoup} and \ref{f.Qunc}).

\begin{figure}[p]
\begin{center}
\includegraphics[width=12cm]{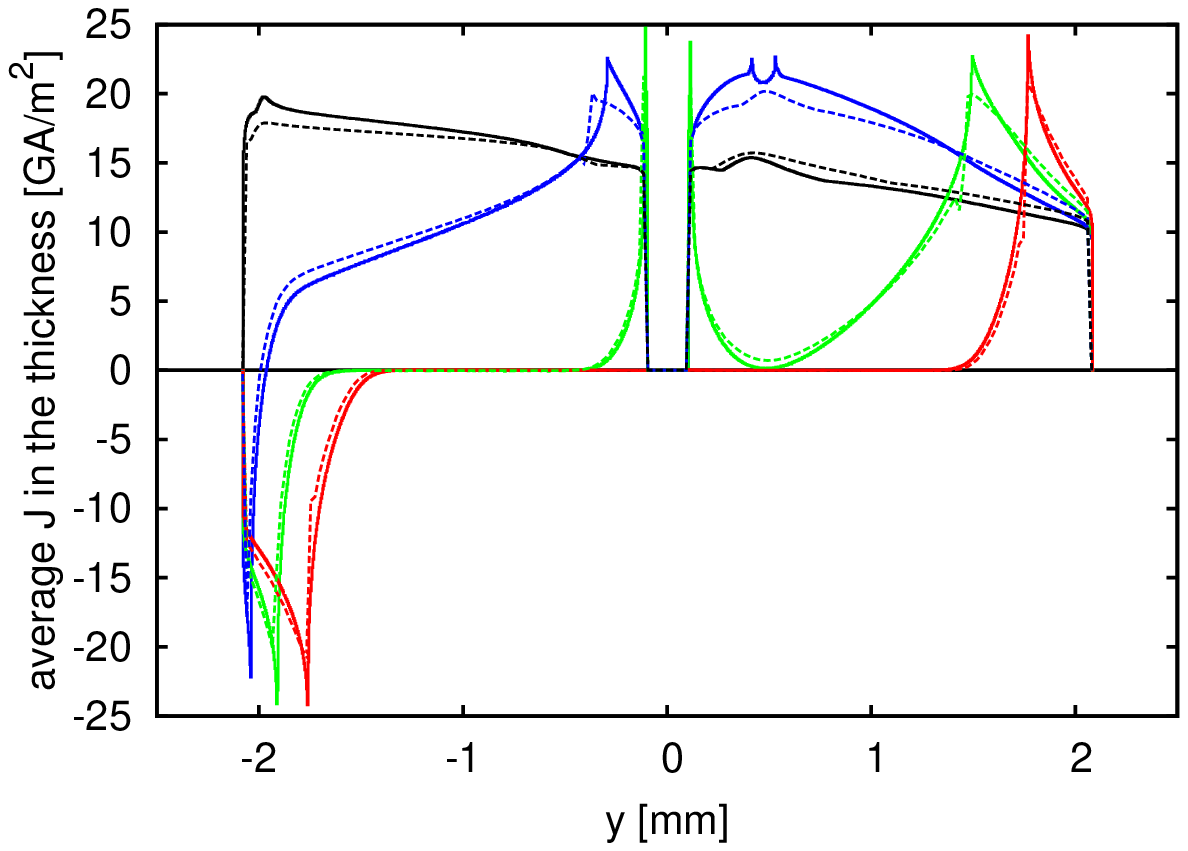}
\includegraphics[width=12cm]{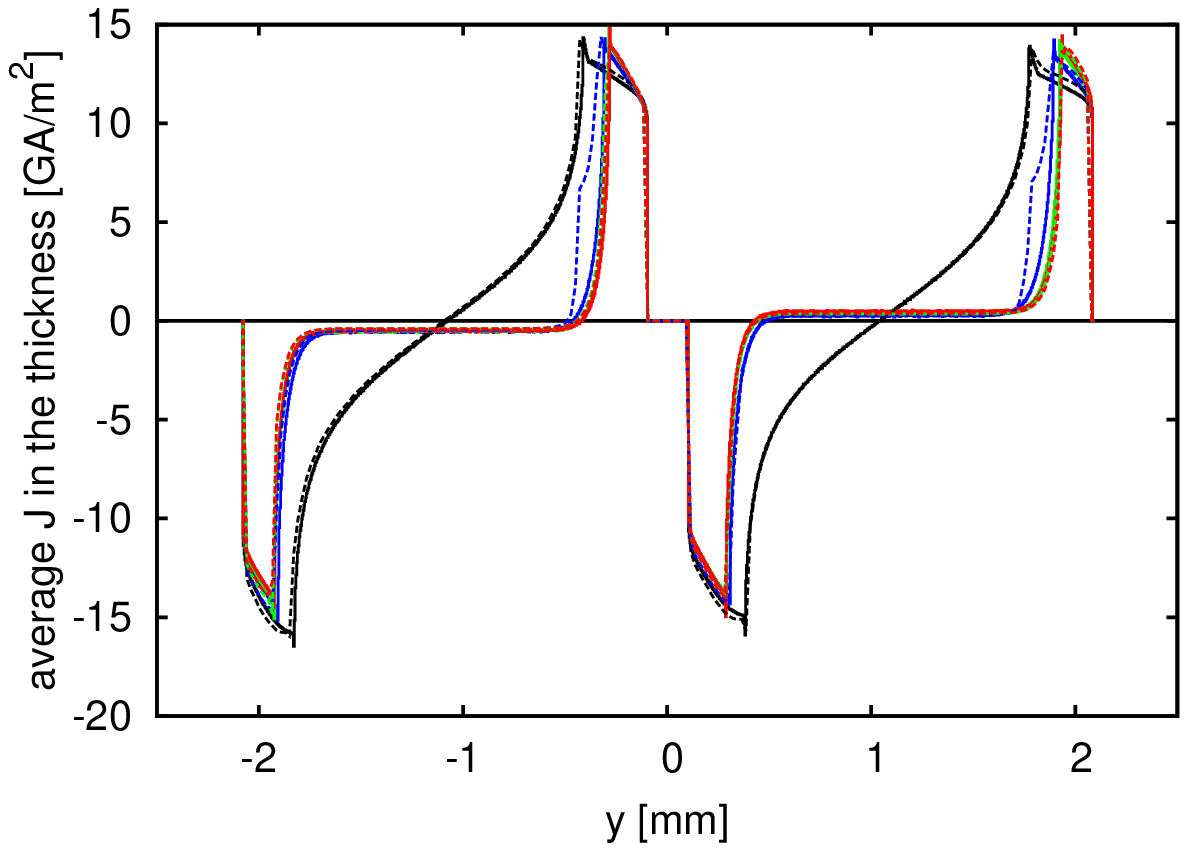}
\end{center}
\caption{The average current density $J$ in the thickness of the tapes for both simulation methods agree (solid and dash lines are for the MMEV and FEM results, respectively). The the different lines (black, blue, green and red) are from the left border to the central layers of tapes in figures \ref{f.Blines} and \ref{f.Jth15}. The situation is the same as in those figures.}\label{f.Kth15}
\end{figure}

The sheet current density, $K$, at the peak of the AC cycle presents the following features (figure \ref{f.Kth15}). Thanks to the inversion point-symmetry, it is enough to study one half of the cable, for example the leftmost half in figure \ref{f.Blines}. For both cases, the sharp peaks in figure \ref{f.Kth15} correspond to the boundary between the regions with sheet current density equal and below its critical value ($K_c=J_cd$). This is because at this boundary $|K|=K_c$ and the magnetic field is minimum (it vanishes for the coupled case, while for the uncoupled case only its perpendicular component vanishes). At the region in between the sharp peaks and the edge of the tape, $|K|$ becomes $K_c$. There, $|K|$ decreases toward the edge because the magnetic field increases. For the FEM simulations, the kinks and small peaks close to the sharp peaks are not necessarily numerical errors. A superconductor with a smooth $E(J)$ relation presents a retarded response comparing to the critical-state model \cite{klincok05SST}. Then, at the peak of the AC field, the maximum current penetration is still not fully developed. For the uncoupled case, the sheet current density around the mid-width of the tapes presents a plateau (figure \ref{f.Kth15}b), as for perpendicular applied fields, transport currents and pancake coils \cite{roebelcomp,pancaketheo}. The plateau is due to the difference in the local magnetic field at both surfaces of the tape and its height does not depend on $J_c$, as detailed in \cite{roebelcomp}. This difference in field appears because of the inhomogeneous field created by the other tapes. Another issue is that the sheet current density close to the lowest $y$-edge of each tape is higher than for the highest $y$-edge. The reason is that at these places the sheet critical current density takes the critical value and the magnetic field concentrates at the top-left and bottom right corners of each stack of tapes, decreasing $J_c$. For the coupled case, the maximum sheet current density is higher because the superconductor shields better the magnetic field, decreasing the field and increasing $J_c$. The sheet current density in each layer of tapes is qualitatively different. In the first layer (black line in figure \ref{f.Kth15}a), the sheet current density roughly decreases with increasing $y$ because the magnetic field increases (figure \ref{f.Blines}). In the second layer (blue line in figure \ref{f.Kth15}a), there is a double peak in the tape at higher $y$ (a double peak for MMEV and single peak for FEM). This is because the magnetic field vanishes in the small region in between the peaks.

When the simulations take into account the penetration into the thickness of the tape, they show that for the uncoupled case and low angles there is an significant penetration in the thickness (figure \ref{f.Jth07}b). Actually, for the situation in figure \ref{f.Jth15}b, the tapes are completely penetrated with current of both signs, with only a small difference corresponding to the height of the plateau in figure \ref{f.Kth15}b. For the coupled case, the distance of penetration across the thickness is much smaller than across the width, therefore the description with a sheet current density is correct in terms of AC loss (note that the actual aspect ratio of the tape is 4000). Moreover, the parallel magnetic field keeps the same sign in the thickness of the tape and, therefore, its average in the thickness is representative for the average critical current density.

\subsection{AC loss for a Roebel cable}

\begin{figure}[p]
\begin{center}
\includegraphics[width=12cm]{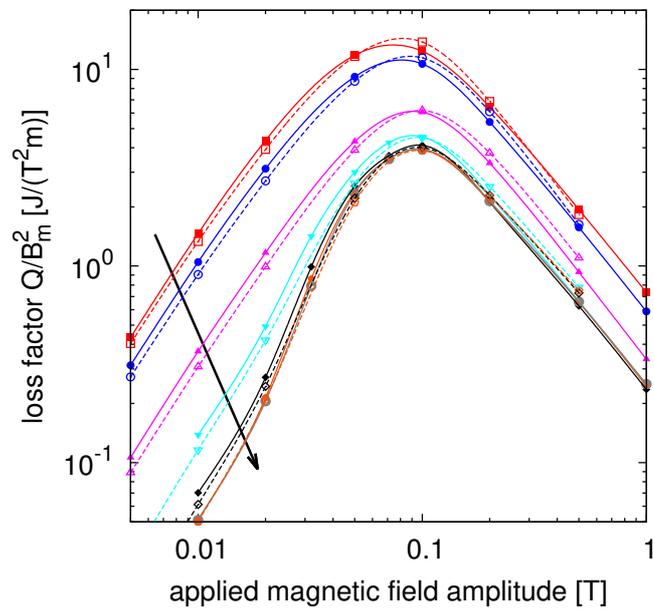}
\end{center}
\caption{Loss factor, $B_m$ is the applied field amplitude, for a Roebel cable in the coupled case. Solid lines with symbols are calculated by MMEV and dash lines with symbols are calculated by FEM. The angle of the applied field amplitude is 90, 60, 30, 15, 7, 3 and 0 degrees in the arrow direction.}\label{f.Qcoup}
\end{figure}

\begin{figure}[p]
\begin{center}
\includegraphics[width=12cm]{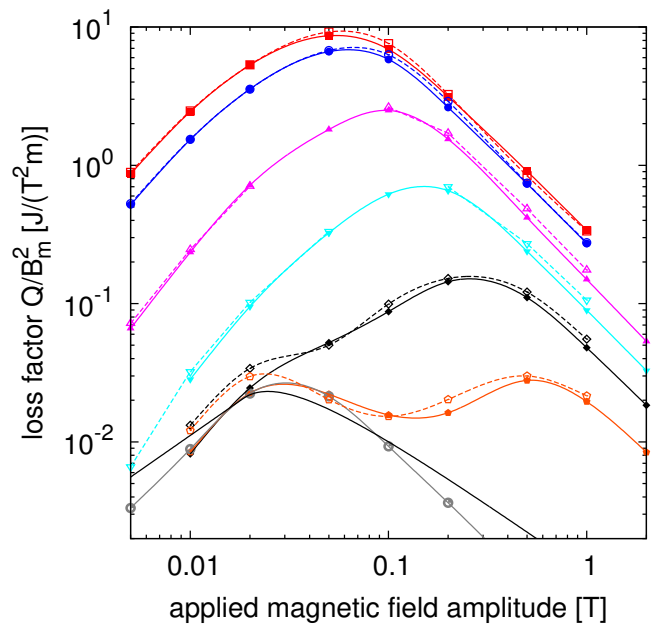}
\end{center}
\caption{Loss factor, $B_m$ is the applied field amplitude, for a Roebel cable in the uncoupled case. Solid lines with symbols are calculated by MMEV and dash lines with symbols are calculated by FEM. The angle of the applied field amplitude is 90, 60, 30, 15, 7, 3 and 0 degrees from top to bottom. The black solid line is for the Bean slab model with a parallel applied magnetic field \cite{goldfarb} (see text).}\label{f.Qunc}
\end{figure}

\begin{figure}[p]
\begin{center}
\includegraphics[width=12cm]{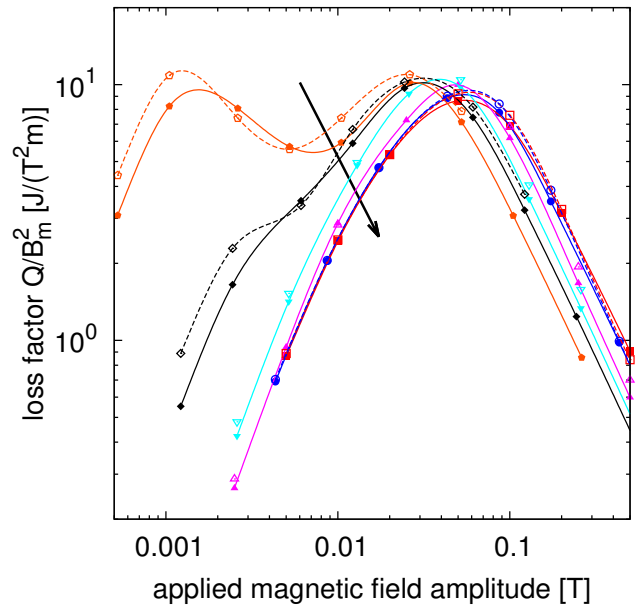}
\end{center}
\caption{The normalised loss factor relative to the perpendicular applied field amplitude, $Q/B_{\rm per}$, for the uncoupled case does not only depend on $B_{\rm per}$ only. Solid lines with symbols are calculated by MMEV and dash lines with symbols are calculated by FEM. The angle of the applied field amplitude is 90, 60, 30, 15, 7 and 3 degrees from top to bottom in the arrow direction.}\label{f.Quncnorm}
\end{figure}

Again, the AC loss calculated from both simulation methods agrees with each other (figures \ref{f.Qcoup}-\ref{f.Quncnorm}). There is a small discrepancy at high amplitudes or low angles and low amplitudes caused by the different physical model of the superconductor (sharp and smooth $E(J)$ relation for MMEV and FEM, respectively). 

The results for the coupled case describe the situation of untransposed tapes. Then, the difference between the AC loss for the coupled and uncoupled cases reveals the maximum possible reduction of AC loss by the transposition in a Roebel cable. For large applied magnetic fields or low angles, the AC loss for the uncoupled case is lower than for the coupled one (figures \ref{f.Qcoup} and \ref{f.Qunc})). The decrease by uncoupling the tapes is especially important for low angles, with a decrease up to three orders of magnitude for parallel fields and high amplitudes. For angles below 15$^{\rm o}$ this reduction is still substantial for all the amplitudes. At high applied fields and angles of 7$^{\rm o}$ or lower the loss reduces more than one order of magnitude compared to the coupled one. 

The AC loss for the coupled case presents the following features (figure \ref{f.Qcoup}). It decreases with decreasing the angle until it saturates to the value for 0$^{\rm o}$. This decrease with decreasing the angle is less pronounced than for a single tape (figure \ref{f.Qtape}). The reason is that the coupled case behaves as a single block \cite{tapesfull,grilli06PhC,roebelcomp} and the aspect ratio of the Roebel cable is relatively small, around 5. Then, the difference in the penetration length across the thickness and the width of the cable is less important than for the tape, resulting in a less pronounced angular dependence.

For the uncoupled case (figure \ref{f.Qunc}), the AC loss is qualitatively similar to a single tape (figure \ref{f.Qtape}). However, there are the following differences. First, the peak in the loss factor is at higher amplitudes because of the stacking effect \cite{tapesfull,roebelcomp,grilli06PhC}. Second, the peak (or peak at higher amplitudes) is wider than for a single tape because of the concentration of the perpendicular field in the gap between the stacks (figure \ref{f.Blines}) \cite{tapesfull,roebelcomp}. Finally, the peak at low applied fields and angles, due to the parallel applied field, becomes more visible. This is because two reasons. First, the peak due to the perpendicular applied field shifts to higher amplitudes, so both peaks do not overlap. Second, the contribution from the parallel field becomes larger {than that of} the perpendicular one. The cause is that the stacking effect reduces the AC loss per volume created by the perpendicular field but the loss created by the parallel field remains the same. For 0$^{\rm o}$, we can compare the simulations with the slab Bean model for a constant $J_c$ \cite{goldfarb,acxrec} taking $J_c$ as the self-field critical current of the tape divided by its volume. The simulations are consistent with those Bean-model calculations (figure \ref{f.Qunc}). The higher and narrower peak of the loss factor for the simulations are due to the field dependence of $J_c$ \cite{chen91JAP}. Then, the Bean model is useful to determine whether the AC loss due to the parallel field is important, assessing the validity of the 1D approximation at low angles without the need of lengthy simulations with several elements in the thickness.

The loss factor relative to the perpendicular applied field $Q/B_{\rm per}^2$ (figure \ref{f.Quncnorm}), does not depend only on $B_{\rm per}$, as is the case for a single tape (figure \ref{f.Qtape}). With decreasing the angle, the shift in the peak and its increase (or the peak at high amplitudes for low angles) is less pronounced than for the single tape. However, for low $B_{\rm per}$, the increase of the loss factor due to the parallel component of the applied field is more important.

In real Roebel cables, the angular dependence of $Q/B_{\rm per}^2$ could be larger. This is because there can be misalignment between the tapes, they can be slightly deformed (like making a circular arch) or the superconductor layer could have a non-negligible roughness -- see for example pictures in~\cite{terzieva11SST}. All these effects will result in larger contributions of the parallel applied field. In addition, coupling currents will also increase the AC loss from the parallel applied field.

The AC loss for low angles will be important for long solenoids, like in some transformers. Therefore, we recommend to characterise the AC loss in Roebel cables and tapes also at low angles of the applied field, therefore avoiding the error committed extrapolating $Q/B_{\rm per}^2$ for low angles assuming an angular independence of $Q/B_{\rm per}^2$.


\section{Summary and conclusions}
\label{s.conclusions}

In this article, we have presented the main features of the magnetic field, current distribution and AC loss for an applied magnetic field with arbitrary orientation, based on two independent numerical simulations. For the Roebel cable, we have taken into account the coupled and uncoupled cases, corresponding to the limits of very high coupling currents and negligible coupling currents, respectively. The simulations for the uncoupled case are also valid for a double pancake coil with no transport current, therefore the results and conclusions for the Roebel cable in the uncoupled case are also applicable to double pancake coils. The simulations have taken into account a realistic anisotropic field dependence of $J_c$. This dependence is in accordance with $J_c$ measurements within around 20\% error. Therefore, the results from the simulations are representative for real tapes, in contrast to those from simple dependences, such as the elliptical or isotropic ones.

About the two simulation methods in this article, we have seen that all the results agree with each other, serving as a mutual check. In addition, the qualitative results for the field and current distributions are well explained by magnetostatic considerations.

For both single tapes and Roebel cables in the coupled case, we found that the AC loss does not only depend on the perpendicular component of the applied field, opposite to the published works \cite{ichiki04PhC,enomoto05IES,amemiya04SSTa,ogawa05Cry,iwakuma04PhC,iwakuma05IES,jiangZ08SST}. For low angles between the applied field and the tapes plane and moderate and large amplitudes, the extrapolated AC loss from purely perpendicular applied fields assuming independence of the parallel component of the applied field is around 3 times lower than the actual one, therefore being this difference not negligible. The discrepancy is even larger for low amplitudes. The AC loss also depends on the parallel component of the applied field for two reasons. First, at low applied fields and low angles, the AC loss due to the penetration across the thickness is important. Second, because for YBCO the field anisotropy in $J_c$ is relatively weak. Then, $J_c$ reduces significantly with increasing the parallel component of the applied field for the same perpendicular component of the applied field. 

For coupled Roebel cables, the AC loss due to the parallel and perpendicular components of the applied field is of the same order of magnitude for all the amplitudes. This is because of the low aspect ratio of the cable, around 5.

Finally, we have found that the highest potential of Roebel cables is to reduce the AC loss at applied fields with low angles with the tapes surface. For angles of 7$^{\rm o}$ or lower, the Robel cable reduces the AC loss more than one order of magnitude, comparing to a cable made of untransposed tapes with the same dimensions. The reduction can be of up to three orders of magnitude for perfectly parallel applied fields. This is a much better reduction than the factor 2 for perpendicular applied fields \cite{roebelmeassim,roebelcomp}.

In conclusion, transposed cables, such as Roebel cables, should be used in windings with an important parallel applied field, such as the low-voltage winding of transformers. For the characterisation of the AC loss of these cables, it is necessary to measure also at low angles with the tape surface ($\le 15$$^{\rm o}$). This consideration is also valid for double (or single) pancake coils to be part of a long solenoid, such as the high-voltage winding of a transformer. As future work, we propose comparison with experiments, the prediction of the influence of intermediate coupling currents and the role of striated tapes in the Roebel cable.

\section*{Acknowledgement}
This work was supported partly by the Structural Funds of the European Union through the Agency for the Structural Funds of the European Union from the Ministry of Education, Science, Research and Sport of the Slovak Republic under the contract number 26240220028 and partly by a Helmholtz-University Young Investigator Grant (VH-NG-617). F. Grilli would like to thank Bertrand Dutoit for the use of the computing resources at Ecole Polytechnique F\'ed\'erale de Lausanne.

\section*{References}
\bibliographystyle{unsrt}	
\bibliography{all20110718.bib}

\begin{thebibliography}{10}

\bibitem{weijers10IES}
H.~W. Weijers, U.~P. Trociewitz, W.~D. Markiewicz, J.~Jiang, D.~Myers, E.~E.
  Hellstrom, A.~Xu, J.~Jaroszynski, P.~Noyes, Y.~Viouchkov, and D.~C.
  Larbalestier.
\newblock High field magnets with {HTS} conductors.
\newblock {\em IEEE Transactions on Applied Superconductivity}, 20(3):576--582,
  2010.

\bibitem{turenne10IPAC}
M.~Turenne, R.~P. Johnson, F.~Hunte, J.~Schwartz, and H.~Song.
\newblock {Roebel} cable for high-field low-loss accelerator magnets.
\newblock {\em Proceedings of the International Particle Accelerator Conference
  (IPAC)}, (MOPEB057), 2010.
\newblock Available at
  http://accelconf.web.cern.ch/accelconf/IPAC10/papers/mopeb057.pdf.

\bibitem{J:2006:Goldacker06}
W.~Goldacker, R.~Nast, G.~Kotzyba, S.~I. Schlachter, A.~Frank, B.~Ringsdorf,
  C.~Schmidt, and P.~Komarek.
\newblock High current {DyBCO-ROEBEL} assembled coated conductor ({RACC}).
\newblock {\em J. Phys.: Conf. Ser.}, 43:901, 2006.

\bibitem{J:2009:Goldacker09}
W.~Goldacker, A.~Frank, A.~Kudymow, R.~Heller, A.~Kling, S.~Terzieva, and
  C.~Schmidt.
\newblock Status of high transport current {ROEBEL} assembled coated conductor
  cables.
\newblock {\em Superconductor Science and Technology}, 22:034003, 2009.

\bibitem{leghissa06AST}
M.~Leghissaa, V.~Hussennetherb, and H.~W. Neum\"uller.
\newblock {kA-}class high-current {HTS} conductors and windings for large scale
  applications.
\newblock {\em Advances in Science and Technology}, 47:212--219, 2006.

\bibitem{J:2008:Long08a}
N.~J. Long, R.~Badcock, P.~Beck, M.~Mulholland, N.~Ross, M.~Staines, H.~Sun,
  J.~Hamilton, and R.~G. Buckley.
\newblock Narrow strand {YBCO} {Roebel} cable for lowered {AC} loss.
\newblock {\em J. Phys.: Conf. Ser.}, 97:012280, 2008.

\bibitem{J:2009:LeeJK09a}
J.~K. Lee, S.~Byun, B.~W. Han, W.~S. Kim, S.~Park, S.~Choi, C.~Park, and
  K.~Choi.
\newblock Reduction effect on magnetization loss in the stacked conductor with
  striated and transposed {YBCO} coated conductor.
\newblock {\em IEEE Transactions on Applied Superconductivity}, 19:3340, 2009.

\bibitem{xieYY09IES}
Y.~Y. Xie, M.~Marchevsky, X.~Zhang, K.~Lenseth, Y.~Chen, X.~Xiong, Y.~Qiao,
  A.~Rar, B.~Gogia, R.~Schmidt, A.~Knoll, V.~Selvamanickam, G.~G. Pethuraja,
  and P.~Dutta.
\newblock Second-generation hts conductor design and engineering for electrical
  power applications.
\newblock {\em IEEE Transactions on Applied Superconductivity},
  19(3):3009--3013, 2009.

\bibitem{mikitik04PRBa}
G.~P. Mikitik, E.~H. Brandt, and M.~Indenbom.
\newblock Superconducting strip in an oblique magnetic field.
\newblock {\em Physical Review B}, 70:014520, 2004.

\bibitem{ichiki04PhC}
Y.~Ichiki and H.~Ohsaki.
\newblock Numerical analysis of {AC} losses in {YBCO} coated conductor in
  external magnetic field.
\newblock {\em Physica C}, 412-414:1015--1020, 2004.

\bibitem{brandt05PRBa}
E.~H. Brandt and G.~P. Mikitik.
\newblock Anisotropic superconducting strip in an oblique magnetic field.
\newblock {\em Physical Review B}, 72:024516, 2005.

\bibitem{enomoto05IES}
N.~Enomoto, T.~Izumi, and N.~Amemiya.
\newblock Electromagnetic field analysis of rectangular superconductor with
  large aspect ratio in arbitrary orientated magnetic fields.
\newblock {\em IEEE Transactions on Applied Superconductivity},
  15(2):1574--1577, 2005.

\bibitem{stavrev05SST}
S.~Stavrev, F.~Grilli, B.~Dutoit, and {S. P.} Ashworth.
\newblock Comparison of the {AC} losses of {BSCCO} and {YBCO} conductors by
  means of numerical analysis.
\newblock {\em Superconductor Science and Technology}, 18(10):1300--1312, 2005.

\bibitem{EUCAS10fmsc}
F.~{G\"om\"ory}, M.~{Vojen\v ciak}, E.~Pardo, M.~Solovyov, and J.~{\v Souc}.
\newblock {AC} losses in coated conductors.
\newblock {\em Superconductor Science and Technology}, 23:034012, 2010.

\bibitem{kiuchi03PhC}
M.~Kiuchi, E.~S. Otabe, T.~Matsushita, T.~Kuga, M.~Inoue, T.~Kiss, Y.~Iijima,
  K.~Kakimoto, and T.~Saitoh.
\newblock Angular dependence of irreversibility field in {Y-123} coated tape.
\newblock {\em Physica C}, 392-396:1063--1067, 2003.

\bibitem{selvamanickam09PhC}
V.~Selvamanickam, Y.~Chen, J.~Xie, Y.~Zhang, A.~Guevara, I.~Kesgin, G.~Majkic,
  and M.~Martchevsky.
\newblock Influence of {Zr} and {Ce} doping on electromagnetic properties of
  {(Gd,Y)$-$Ba$-$Cu$-$O} superconducting tapes fabricated by metal organic
  chemical vapor deposition.
\newblock {\em Physica C}, 469:2037--2043, 2009.

\bibitem{holesinger09SST}
T.~G. Holesinger, B.~Maiorov, O.~Ugurlu, L.~Civale, Y.~Chen, X.~Xiong, Y.~Xie,
  and V.~Selvamanickam.
\newblock Microstructural and superconducting properties of high current
  metal$-$organic chemical vapor deposition {YBa$_2$Cu$_3$O$_{7-\delta}$}
  coated conductor wires.
\newblock {\em Superconductor Science and Technology}, 22:045025, 2009.

\bibitem{zhangY09PhC}
Y.~Zhang, E.D. Specht, C.~Cantoni, D.K. Christen, J.R. Thompson, J.W. Sinclair,
  A.~Goyal, Y.L. Zuev, T.~Aytug, M.P. Paranthaman, Y.~Chen, and
  V.~Selvamanickam.
\newblock Magnetic field orientation dependence of flux pinning in
  {(Gd,Y)Ba$_2$Cu$_3$O$_{7-x}$} coated conductor with tilted lattice and
  nanostructures.
\newblock {\em Physica C}, 469:2044--2051, 2009.

\bibitem{CoatedIc}
E.~Pardo, M.~{Vojen\v ciak}, F.~{G\"om\"ory}, and J.~{\v Souc}.
\newblock Low-magnetic-field dependence and anisotropy of the critical current
  density in coated conductors.
\newblock {\em Superconductor Science and Technology}, 24:065007, 2011.

\bibitem{roebelmeassim}
S.~Terzieva, M.~{Vojen\v ciak}, E.~Pardo, F.~Grilli, A.~Drechsler, A.~Kling,
  A.~Kudymow, F.~{G\"om\"ory}, and W.~Goldacker.
\newblock Transport and magnetization ac losses of {ROEBEL} assembled coated
  conductor cables: measurements and calculations.
\newblock {\em Superconductor Science and Technology}, 23:014023, 2010.

\bibitem{roebelcomp}
F.~Grilli and E.~Pardo.
\newblock Simulation of ac loss in roebel coated conductor cables.
\newblock {\em Superconductor Science and Technology}, 23:115018, 2010.

\bibitem{jiangZ11SST}
Z.~Jiang, K.~P. Thakur, M.~Staines, R.~A. Badcock, N.~J. Long, R.~G. Buckley,
  A.~D. Caplin, and N.~Amemiya.
\newblock The dependence of ac loss characteristics on the spacing between
  strands in ybco roebel cables.
\newblock {\em Superconductor Science and Technology}, 24:065005, 2011.

\bibitem{tapesfull}
E.~Pardo, A.~Sanchez, and C.~Navau.
\newblock Magnetic properties of arrays of superconducting strips in a
  perpendicular field.
\newblock {\em Physical Review B}, 67:104517, 2003.

\bibitem{grilli06PhC}
F.~Grilli, S.~P. Ashworth, and S.~Stavrev.
\newblock Magnetization ac losses of stacks of {YBCO} coated conductors.
\newblock {\em Physica C}, 434:185--190, 2006.

\bibitem{yuanW10JAP}
W.~Yuan, A.~M. Campbell, and T.~A. Coombs.
\newblock {ac} losses and field and current density distribution during a full
  cycle of a stack of superconducting tapes.
\newblock {\em Journal of Applied Physics}, 107:093909, 2010.

\bibitem{prigozhin11SST}
L.~Prigozhin and V.~Sokolovsky.
\newblock Computing ac losses in stacks of high-temperature superconducting
  tapes.
\newblock {\em Superconductor Science and Technology}, 24:075012, 2011.

\bibitem{terzieva11SST}
S.~Terzieva, M.~{Vojen\v ciak}, F.~Grilli, R.~Nast, J.~{\v Souc}, W.~Goldacker,
  A.~Jung, A.~Kudymow, and A.~Kling.
\newblock Investigation of the effect of striated strands on the {AC} losses of
  {2G Roebel} cables.
\newblock {\em Superconductor Science and Technology}, 24:045001, 2011.

\bibitem{amemiya04SSTa}
N.~Amemiya, T.~Nishioka, Z.~Jiang, and K.~Yasuda.
\newblock Influence of film width and magnetic field orientation on ac loss in
  ybco thin film.
\newblock {\em Superconductor Science and Technology}, 17:485--492, 2004.

\bibitem{ogawa05Cry}
J.~Ogawa, H.~Nakayama, S.~Odaka, and O.~Tsukamoto.
\newblock {AC} loss characteristics of {YBCO} conductors carrying transport
  currents in external {AC} magnetic fields.
\newblock {\em Cryogenics}, 45:23--27, 2005.

\bibitem{iwakuma04PhC}
M.~Iwakuma, K.~Toyota, M.~Nigo, T.~Kiss, K.~Funaki, Y.~Iijima, T.~Saitoh,
  Y.~Yamada, and Y.~Shiohara.
\newblock {AC} loss properties of {YBCO} superconducting tapes fabricated by
  {IBAD-PLD} technique.
\newblock {\em Physica C}, 412-414:983--991, 2004.

\bibitem{iwakuma05IES}
M.~Iwakuma, M.~Nigo, D.~Inoue, T.~Kiss, K.~Funaki, Y.~Iijima, T.~Saitoh,
  Y.~Yamada, and Y.~Shiohara.
\newblock {AC} loss properties of {YBCO} superconducting tapes exposed to
  external ac magnetic field.
\newblock {\em IEEE Transactions on Applied Superconductivity},
  15(2):1562--1565, 2005.

\bibitem{jiangZ08SST}
Z.~Jiang, N.~Amemiya, K.~Kakimoto, Y.~Iijima, T.~Saitoh, and Y.~Shiohara.
\newblock The dependence of {AC} loss characteristics on the space in stacked
  {YBCO} conductors.
\newblock {\em Superconductor Science and Technology}, 21:015020, 2008.

\bibitem{lakshmi10SSTa}
L.~S. Lakshmi, M.~P. Staines, K.~P. Thakur, R.~A. Badcock, and N.~J. Long.
\newblock Frequency dependence of magnetic ac loss in a five strand {YBCO}
  roebel cable.
\newblock {\em Superconductor Science and Technology}, 23:065008, 2010.

\bibitem{J:2010:lakshmi10b}
L.~S. Lakshmi, M.~P. Staines, R.~A. Badcock, N.~J. Long, M.~Majoros, E.~W.
  Collings, and M.~D. Sumption.
\newblock Frequency dependence of magnetic ac loss in a roebel cable made of
  {YBCO} on a {Ni–W} substrate.
\newblock {\em Superconductor Science and Technology}, 23:085009, 2010.

\bibitem{J:2010:jiangZ10}
Z.~Jiang, R.~A. Badcock, N.~J. Long, M.~Staines, K.~P. Thakur, L.~S. Lakshmi,
  A.~Wright, K.~Hamilton, G.~N. Sidorov, R.~G. Buckley, N.~Amemiya, and A.~D.
  Caplin.
\newblock Transport ac loss characteristics of a nine strand {YBCO} roebel
  cable.
\newblock {\em Superconductor Science and Technology}, 23:025028, 2010.

\bibitem{pancaketheo}
E.~Pardo.
\newblock Modeling of coated conductor pancake coils with a large number of
  turns.
\newblock {\em Superconductor Science and Technology}, 21:065014, 2008.

\bibitem{simHTS11}
E.~Pardo, J.~{\v Souc}, F.~{G\"om\"ory}, and F.~Grilli.
\newblock {AC} loss in stacks of pancake coils made of coated conductor:
  simulations agree with measurements.
\newblock {\em 2nd International Workshop on Numerical Modelling of High
  Temperature Superconductors}, 2011.
\newblock Available at
  http://www-g.eng.cam.ac.uk/epec/HTSModellingWorkgroup/program/index.php.

\bibitem{S:COMSOL}
Finite-element software package {Comsol Multiphysics}. http://www.comsol.com.

\bibitem{nguyen10SST}
D.~N. Nguyen, S.~P. Ashworth, J.~O. Willis, F.~Sirois, and F.~Grilli.
\newblock A new finite-element method simulation model for computing ac loss in
  roll assisted biaxially textured substrate ybco tapes.
\newblock {\em Superconductor Science and Technology}, 23:025001, 2010.

\bibitem{roy08IES}
F.~Roy, B.~Dutoit, F.~Grilli, and F.~Sirois.
\newblock Magneto-thermal modeling of 2nd generation {HTS} for resistive fault
  current limiter design purposes.
\newblock {\em IEEE Transactions on Applied Superconductivity}, 18(1):29--35,
  2008.

\bibitem{prigozhin97IES}
L.~Prigozhin.
\newblock Analysis of critical-state problems in type{-II} superconductivity.
\newblock {\em IEEE Transactions on Applied Superconductivity},
  7(4):3866--3873, 1997.

\bibitem{sanchez01PRB}
A.~Sanchez and C.~Navau.
\newblock Magnetic properties of finite superconducting cylinders. i. uniform
  applied field.
\newblock {\em Physical Review B}, 64:214506, 2001.

\bibitem{HacIacinphase}
E.~Pardo, F.~{G\"om\"ory}, J.~{\v Souc}, and J.~M. Ceballos.
\newblock Current distribution and ac loss for a superconducting rectangular
  strip with in-phase alternating current and applied field.
\newblock {\em Superconductor Science and Technology}, 20(4):351--364, 2007.

\bibitem{pancakeBi}
J.~{\v Souc}, E.~Pardo, M.~{Vojen\v ciak}, and F.~{G\"om\"ory}.
\newblock Theoretical and experimental study of ac loss in high temperature
  superconductor single pancake coils.
\newblock {\em Superconductor Science and Technology}, 22:015006, 2009.

\bibitem{pancakeFM}
E.~Pardo, J.~{\v Souc}, and M.~{Vojen\v ciak}.
\newblock {AC} loss measurement and simulation of a coated conductor pancake
  coil with ferromagnetic parts.
\newblock {\em Superconductor Science and Technology}, 22:075007, 2009.

\bibitem{brambilla07SST}
R.~Brambilla, F.~Grilli, and L.~Martini.
\newblock Development of an edge-element model for {AC} loss computation of
  high-temperature superconductors.
\newblock {\em Superconductor Science and Technology}, 20(1):16--24, 2007.

\bibitem{SuperPower}
SuperPower, Inc. http://www.superpower-inc.com/.

\bibitem{polak09SST}
M.~Polak, S.~Takacs, P.~N. Barnes, and G.~A. Levin.
\newblock The effect of resistive filament interconnections on coupling losses
  in filamentary {YBa$_2$Cu$_3$O$_7$} coated conductors.
\newblock {\em Superconductor Science and Technology}, 22:034003, 2009.

\bibitem{blatter94RMP}
G.~Blatter, M.~V. Feigelman, V.~B. Geshkenbein, A.~I. Larkin, and V.~M.
  Vinokur.
\newblock Vortices in high-temperature superconductors.
\newblock {\em Rev. Mod. Phys.}, 66:1125, 1994.

\bibitem{oomen97APL}
M.~P. Oomen, J.~Rieger, M.~Leghissaa, and H.~H.~J. {ten Kate}.
\newblock Field-angle dependence of alternating current loss in
  multifilamentary high{-$T_c$} superconducting tapes.
\newblock {\em Applied Physics Letters}, 70:3038--3040, 1997.

\bibitem{chen91JAP}
D.~X. Chen and A.~Sanchez.
\newblock Theoretical critical-state susceptibility spectra and their
  application to high-{$T_c$} superconductors.
\newblock {\em Journal of Applied Physics}, 70(10):5463--5477, 1991.

\bibitem{klincok05SST}
B.~{Klin\v cok}, F.~{G\"om\"ory}, and E.~Pardo.
\newblock The voltage signal on a superconducting wire in ac transport.
\newblock {\em Superconductor Science and Technology}, 18:694--700, 2005.

\bibitem{goldfarb}
R.~B. Goldfarb, M.~Lelental, and C.~Thompson{, edited by R. A. Hein, p. 49}.
\newblock {\em Magnetic Susceptibility of Superconductors and Other Spin
  Systems}.
\newblock Plenum, New York, 1991.

\bibitem{acxrec}
E.~Pardo, D.-X. Chen, A.~Sanchez, and C.~Navau.
\newblock The transverse critical-state susceptibility of rectangular bars.
\newblock {\em Supercond. Sci. Technol.}, 17:537, 2004.

\end{thebibliography}


\newpage
\newpage

\begin{figure}[p]
\begin{center}
\includegraphics[width=12cm]{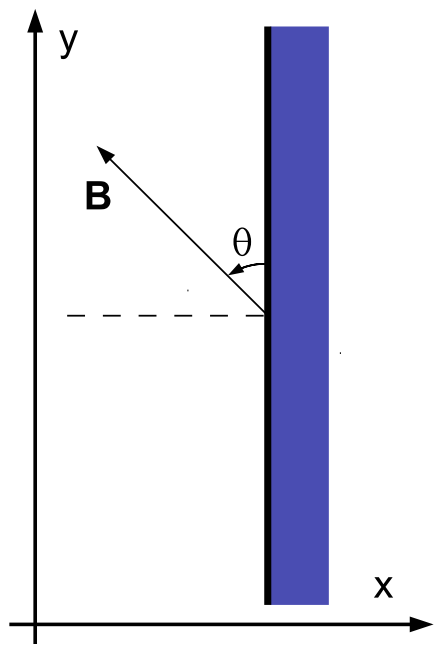}
\end{center}
\caption{The sketch shows the origin of the angle $\theta$ between magnetic field $\bf B$ and the surface of the tape.}\label{f.sketch}
\end{figure}

\newpage

\end{document}